\documentclass[aps,pre,floatfix,10pt]{revtex4}
\usepackage{graphicx}
\usepackage{dcolumn}
\usepackage{bm}
\usepackage{latexsym}
\usepackage{amsmath}
\usepackage{amssymb}
\usepackage{color}
\usepackage{ulem}

\begin{document}
\title{{\bf Flagellar length control in monoflagellates by motorized transport:\\ growth kinetics and correlations of length fluctuations}}
\author{Swayamshree Patra} 
\affiliation{Department of Physics, Indian
  Institute of Technology Kanpur, 208016, India} 
\author{Debashish Chowdhury{\footnote{E-mail: debch@iitk.ac.in}}}
\affiliation{Department of Physics, Indian Institute of Technology
  Kanpur, 208016, India}

\begin{abstract}
 How does a cell self-organize so that its appendages attain specific lengths that are convenient for their respective functions? What kind of `rulers' does a cell use to measure the length of these appendages? How does a cell transport structure building materials between the cell body and distal tips of these appendages so as to regulate their dynamic lengths during various stages of its lifetime? Some of these questions are addressed here in the context of a specific cell appendage called flagellum (also called cilium). A ``time of flight'' (ToF) mechanism, adapted from the pioneering idea of Galileo, has been used successfully very recently to explain the length control of flagella by a biflagellate green algae. Using the same ToF mechanism, here we develop a stochastic model for the dynamics of flagella in two different types of monoflagellate unicellular organisms. A unique feature of these monoflagellates is that these become transiently multi-flagellated during a short span of their life time. The mean length of the flagella in our model reproduce the trend of their temporal variation observed in experiments. Moreover, for probing the intracellular molecular communication between the dynamic flagella of a given cell, we have computed the correlation in the fluctuations of their lengths during the multiflagellated stage of the cell cycle by Monte Carlo simulation.   
\end{abstract}

\maketitle

\section{Dedication}

Dietrich Stauffer (DS) was the external examiner of the Ph.D. thesis of the senior author (DC) of this paper in 1984.  DS was also his first post-doc mentor (host professor during DC's visit to Cologne as an Alexander von Humboldt fellow) and then a long-time collaborator. DC had the privilege of writing more than 30 papers and a book with DS over the twenty year period 1985-2005. DS deserved to be co-author of several other papers of DC, but DS refused to put his name on those papers because he had not himself computed any of the results reported in those despite supplying some of the key ideas. Outside science, DS and DC shared a common interest in history on which  their discussion and debate continued for thirty-five years till January 2019 although their views did not necessarily converge on the causes and consequences of many historical events. This paper is dedicated to the memory of DS with whom DC wrote his first paper on theoretical biology.

\section{Introduction}

Cell is the structural and functional unit of a living organism. Cilia are hair like appendages that protrude outward from the membrane of eukaryotic cells \cite{lee18}. These are also referred to as flagella; although we use the terms cilia and flagella interchangeably, these should not be confused with the bacterial flagella.  In the fully grown stage, a particular type of cilia in a fixed cell type of a specific organism has a length that varies very little from one member of the organism to another. There is near unanimity among biologists that these cell protrusions have the right size for the biological functions they perform \cite{marshall15b}. But, what is still hotly debated are the following questions: (i) what are  the `rulers' that  a cell uses to measure the lengths of its protrusions like flagella \cite{marshall15a,marshall15b}, and (ii) given a `ruler', how does a cell normally assemble, regulate and disassemble flagella during various stages of its life, called cell cycle, and also in the face of external stress \cite{ludington15,patra20}? In this paper we address the question (ii) assuming the validity of a scenario that has been discussed widely in the literature as a possible answer to the question (i). 

The `ruler' used by some unicellular organisms for measuring flagellar length is actually a `timer' \cite{ishikawa17,webb16}. The time-of-flight (ToF) of a timer molecule in a single round trip, from the base to the tip of the flagellum and return to the base, provides a measure of the flagellar length; for a given velocity of travel,  the longer is the flagellum the longer is the ToF. The basic principle of ToF mechanism, pioneered by Galileo, is widely used for distance measurements using suitable signals like light or sound, etc. Very recently, we have developed a quantitative model of flagellar length control where the cell senses the length of a flagellum from the feedback it receives from the ToF of a timer molecule \cite{patra20}. The same ToF mechanism is adopted in this paper for modeling dynamics of flagellar lengths in monoflagellates, i.e., cells that possess a single flagellum. 

The phenomenon of assembling cilia is known as ciliogenesis\cite{lechtreck17,ishikawa17}. The machineries for synthesizing the materials  for the structure of the cilia (from now onwards, we refer to these protein molecules as ``precursor'') exist only in the cell body and not in the ciliary protrusions. Besides, the cilia grow by adding new subunits at its distal end and not at its base. The transport of the fresh supply of precursors from the cell body to the distal tip and that of the degraded precursors from the tip to the base is a challenging logistical problem for a flagellated cell. It solves this problem of bidirectional transport  with an elaborate system known as intraflagellar transport (IFT) \cite{lechtreck17}. 

Traffic over wide range of scales, starting from macroscopic vehicular traffic on highways to molecular motor \cite{chowdhury13,kolomeisky15} traffic in living cells \cite{chowdhury05,chowdhury00,schad10}, have been modelled by suitable extensions of the totally asymmetric simple exclusion process (TASEP) \cite{parmeggiani03,nishinari05,greulich07,chowdhury08b,zia11,chou11,patra20,sugden07,schmitt11,rolland15,johann12,melbinger12,ghosh18,sharma12,pinkoviezky13,sahoo16,patra18,john09}.
As is well known, TASEP is one of the simplest models of systems of interacting self-driven particles that can support steady flow of the particles in its non-equilibrium steady states. The models that we develop in this paper incorporate IFT by suitably adapting the formalism of TASEP. 

So far ciliogenesis has been investigated mostly using the biflagellated green algae {\it Chlamydomonas reinhardtii} ({\it C. reinhardtii}) \cite{rosenbaum69,marshall01}. There is strong experimental evidence that the dynamics of the two flagella of {\it C. reinhardtii} cell are coupled at their base through the shared pool of precursors \cite{ludington12}.  Developing a theoretical model that accounts for the cooperativity in the dynamics of flagellar length of biflagellates is highly nontrivial although significant progress have been made. The correlation between the fluctuations of the lengths of the two flagella of {\it C. reinhardtii} displays their cooperativity mediated by the shared common precursor pool \cite{patra20,bauer20}. So, naively, one would expect the dynamics of monoflagellates to be simpler than those in biflagellates \citep{heimann89,bertiaux18}. But, surprisingly, dynamics of monoflagellates have not received as much attention as the biflagellates. 

A possible reason for this lack of interest in the monoflagellates is that these become transiently multi-flagellated during a short span of their life time; the sequence of all the stages of its life, from birth to its ultimate division into two daughter cells, is collectively known as the cell cycle. Each of these monoflagellate cells has a single flagellum during the longest phase of its life called interphase. But, just before the process of cell division begins it synthesizes new flagellum thereby becoming transiently multiflagellated. During this relatively short period prior to cell division the length of one or more of the flagella change with time. Soon after cell division both the nascent daughter cells attain their monoflagellate status. In contrast, in the biflagellate CR, both the flagella of the mother cell are completely resorbed before cell division begins and neither of the daughter cells possess any flagellum at birth; then they synthesize their own flagella which grow during the process of ciliogenesis \cite{heimann89}. This paper is an attempt to draw the attention of physicists to the  challenging problem of understanding ciliary dynamics of monoflagellates by reporting our theoretical models and results for these systems.

\section{Ciliogenesis and cell cycle in monoflagellates: a brief summary}
\label{sec-organisms}

In contrast to {\it C. reinhardtii}, in which the cell division, cilogenesis and flagellar resorprtion occurs sequentially at distinct interval of time, these three processes overlap in the monoflagellates \cite{heimann89}. During some phases of the cell cycle, particularly just before cell division, a monoflagellate actually possesses multiple flagella although in the interphase it is a strictly monoflagellate.  How do the multiple flagella coordinate among themselves when the cell is in a multiflagellated state? How does the ongoing process of redistribution of flagellar precursors during cell division affect the growth kinetics of the flagella? We address  these questions by analyzing theoretical models that quantify two qualitative scenarios that were described earlier in the experimental literature \cite{heimann89,he19,bertiaux18}.  These models account for the  interplay of ciliary growth and shrinkage, precursor population dynamics and cell division.

\begin{figure}
\includegraphics[scale=1.05]{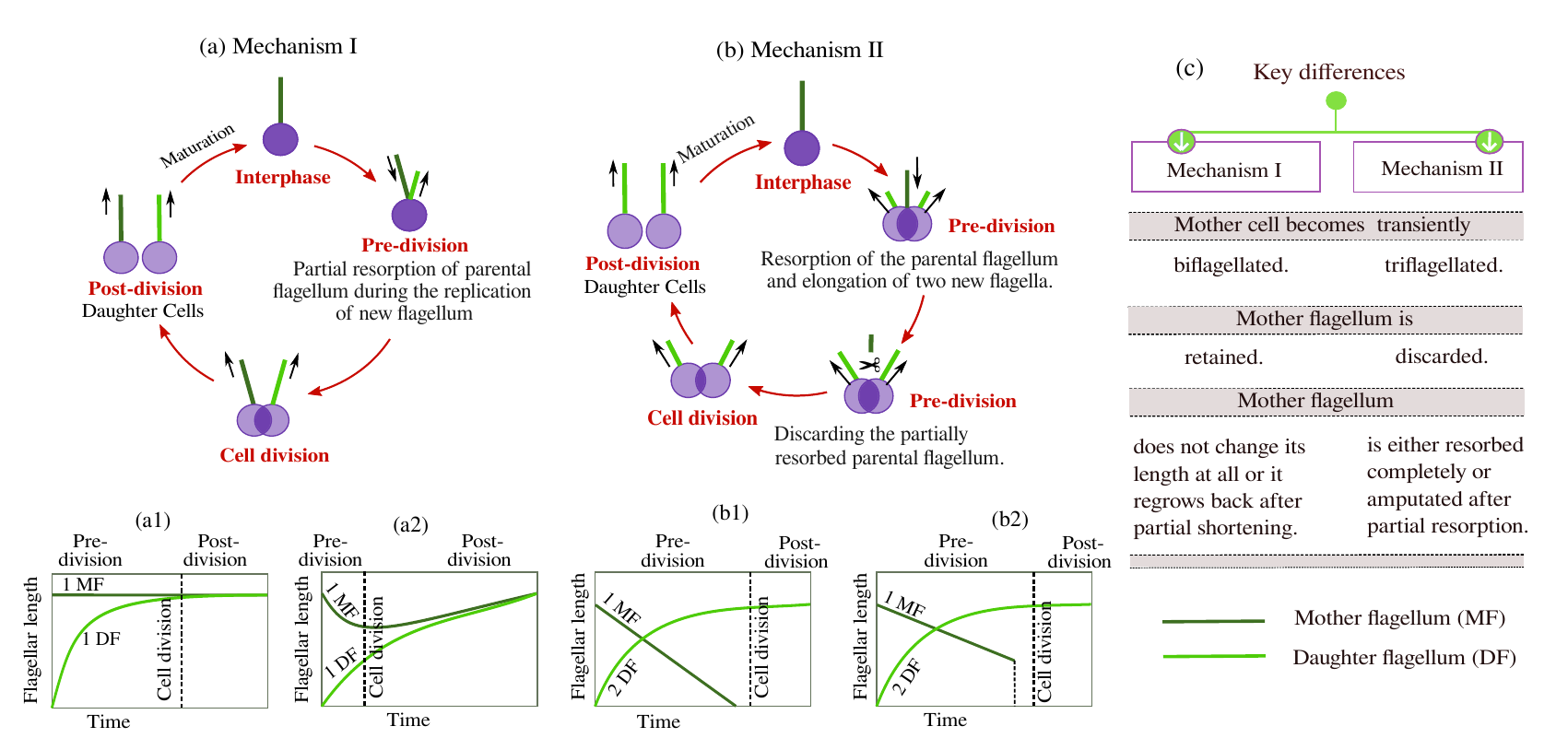}
\caption{ \textbf{Ciliogenesis and cell division in monoflagellates:} Schematic depiction of flagellar growth during cell division in monoflagellates by (a) mechanism I and (b) mechanism II. Temporal evolution of the length  of the old mother flagellum (MF) and the new daughter flagellum (DF) in the uniflagellated cells in which ciliogenesis and cell division take place by mechanism I (a1-a2) and mechanism II (b1-b2). In a1-a2 and b1-b2, 1 MF indicates a single old mother flagellum while 1 DF and 2 DF indicate one new daughter flagellum and two new daughter flagella respectively. (c) List of key differences between mechanisms I and II are listed. } 
\label{mono-all}
\end{figure}

\subsection{Mechanism I}

Flagellar dynamics of {\textit{Pedinomonas tuberculata}}  ({\it P. tuberculata}), a monoflagellate cell, is a physical realization of what we refer to as the mechanism I. It has a small cell body of dimension 4-7$\mu$m and with a flagellum of length 8.8$\pm$0.89 $\mu$m in the interphase \cite{heimann89}. Prior to cell division, a  new  daughter flagellum is synthesized. During this phase, the length of the pre-existing mother  flagellum remains unchanged while the new flagellum grows till its length becomes equal to that  of the existing flagellum.  In the subsequent phase of cell division, there is no alteration in the length of either of the flagella. The pre-existing mother flagellum and the new daughter flagellum synthesized just before cell division are inherited by the two daughter cells, one flagellum each. 

Another monoflagellate {\it Leptomonas pyrrhocoris} ({\it L. pyrrhocoris}) bears a single flagellum of length $20-25 ~ \mu$m \cite{he19}. Prior to cell division, it also assembles an additional flagellum. During the elongation of the new daughter flagellum the existing mother flagellum keeps shortening with time till the cell division starts.  One of the daughter cell inherits this old shortening flagellum whereas the other inherits the new elongating flagellum. After the cell division, both the flagella of the new daughter cells keep elongating till they achieve their matured length. Note that the length of the new elongating flagellum never exceeds that of the old shortening flagellum. 

We summarize the mechanism I observed in the monoflagellates  in Fig.\ref{mono-all}(a). When the replication of the new daughter flagellum starts, the mother cell initiates resorbing its only pre-existing flagellum (the mother flagellum) which has the steady length  $L^{\text{(In)}}_\text{M}$ during the interphase. The minimum length the mother flagellum attains during this resorption phase is $(1-f_\text{MIN} )L^{\text{(In)}}_\text{M}$ ($0 \leq f_\text{MIN} \leq 1$). When the length of the new daughter flagellum becomes  $f_\text{CD} L^{\text{(In)}}_\text{M}$ ($0 \leq f_\text{CD} \leq 1)$, the cell divides into two new daughter cells and each cell inherits a flagellum. If not grown to their full interphase length prior to the cell division, the flagella keep growing till they achieve  their full length. Note that,  $f_\text{CD}=1$ and $f_\text{MIN}=0$ for {\it P. tuberculata} and $0<f_\text{CD}<(1-f_\text{MIN})<1$ for {\it L. pyrrhocoris}. Hence, the generalised version of mechanism I is obseved when $0<f_\text{CD}<1$ and $0<f_\text{MIN}<1$  (for example, for {\it L. pyrrhocoris}) and  we get the special case with $f_\text{MIN}=0$ $f_\text{CD}=1$ (for {\it P. tuberculata}). In Fig.\ref{mono-all}(a) we have pictorially depicted the main stages of the cell cycle and the dynamics of the flagella of the monoflagellated cell in which ciliogenesis progress through mechanism I. The temporal evolution of the old mother flagellum  and the new daughter flagellum is schematically depicted for the {\it P. tuberculata} cell in Fig.\ref{mono-all}(a1) and for the {\it L. pyrrhocoris} cell in Fig.\ref{mono-all}(a2).

\subsection{Mechanism II}

What we refer to as the mechanism II for flagellar kinetics in monoflagellates is physically realized in  {\textit{Pseudopedinella elastica }} ({\it P. elastica }) \cite{heimann89}. It  possess a 20 $\mu$m long flagellum in the interphase. About 6 hours prior to the cell division the flagellum starts to shorten and gets completely resorbed by 15-30 minutes before cell division. But, even before its complete resorption, by the time it shortens to a length of about 2 $\mu$m, two new flagella  emerge almost simultaneously from the cell body. Before the mother cell divides into two daughter cells, the two new flagella grow to about 10 $\mu$m, which is approximately half of their maximum interphase length. After cell division, the flagella inherited by the two daughter cells grow till they attain their normal interphase length 20 $\mu$m.

In another monoflagellate {\textit{Monomastix spec}}  the flagellum, which in the interphase has a length of about 6 $\mu$m, also starts to shorten prior to cell division \cite{heimann89}. But when it shortens to about 3-4 $\mu$m,  two new flagella emerge and start growing. Afterwards, the shortening mother flagellum is discarded completely by a process called deflagellation. The two newly synthesized flagella are distributed among the two daughter cells and the cell division is completed within 5 minutes. The flagella, which are only 3 $\mu$m long in the nascent daughter cells, continue to grow till attaining their normal interphase length 6 $\mu$m. 

We summarize the second mechanism used by the monoflagellates  in Fig.\ref{mono-all}(b). The mother cell initiates resorbing its only flagellum (mother flagellum) which has length  $L^{\text{(In)}}_\text{M}$ during the interphase. When the mother flagellum shortens to length $f_\text{NF} L^{\text{(In)}}_\text{M}$ ($0<f_\text{NF}<1$), two new daughter flagella emerge from the daughter pools. The mother flagellum  further shortens to length $f_\text{A} L^{\text{(In)}}_\text{M}$ ($0<f_\text{A}<f_\text{NF}$). The cell gets rid of the mother flagellum either by complete resorprtion ($f_\text{A} = 0$) or by deflagellation after partial resorption ($f_\text{A} \neq 0$). Following the subsequent division of the mother cell,  the daughter cells inherit one of the two new daughter flagella. The nascent flagella in the new born daughter cells are shorter than their normal interphase length and, therefore, continue to grow till attaining that length. The general version of the mechanism II with $f_A \neq 0$ corresponds to the flagellar dynamics in ({\it Monomastix spec}) while its special case for $f_A=0$  captures that in ({\it P. elastica}). The temporal evolution of the old mother flagellum and the pair of new  daughter flagella are schematically depicted for the {\it P. elastica} cell in Fig.\ref{mono-all}(b1) and for the {\it Monomastix spec} cell in Fig.\ref{mono-all}(b2). 

Let us emphasize the key differences between the two mechanisms (Fig.\ref{mono-all}(c)): (i) In mechanism I, for a brief period prior to cell division the cell becomes {\it biflagellated}. In this mechanism, the mother flagellum is not discarded; instead, it is inherited as the flagellum of one of the two daughter cells while the other daughter takes possession of the new flagellum synthesized by the mother just before cell division. (ii) In mechanism II, for a relatively short duration prior to cell division, the cell appears {\it triflagellated}. Moreover, removal of the flagellum of the mother cell is  completed before cell divides into two daughter cells each of which inherits one of the two new flagella that the mother cell synthesizes just before her division. 
                                                                                                                                                                                                                                                                                                                                                                                                                                                                                                                                                       
\section{Modeling kinetics of flagellar length and the precursor pool} 
\label{sec-model}
The flagellum and precursor pool are two different functional modules of the cell but strongly dependent on each other. In this section, we begin by formulating the dynamical equations that govern (a) the time-dependence of the length of a single flagellum and (b) the population kinetics of the precursors in the pool in the interphase of a monoflagellate. 

\begin{figure}
\includegraphics[scale=0.8]{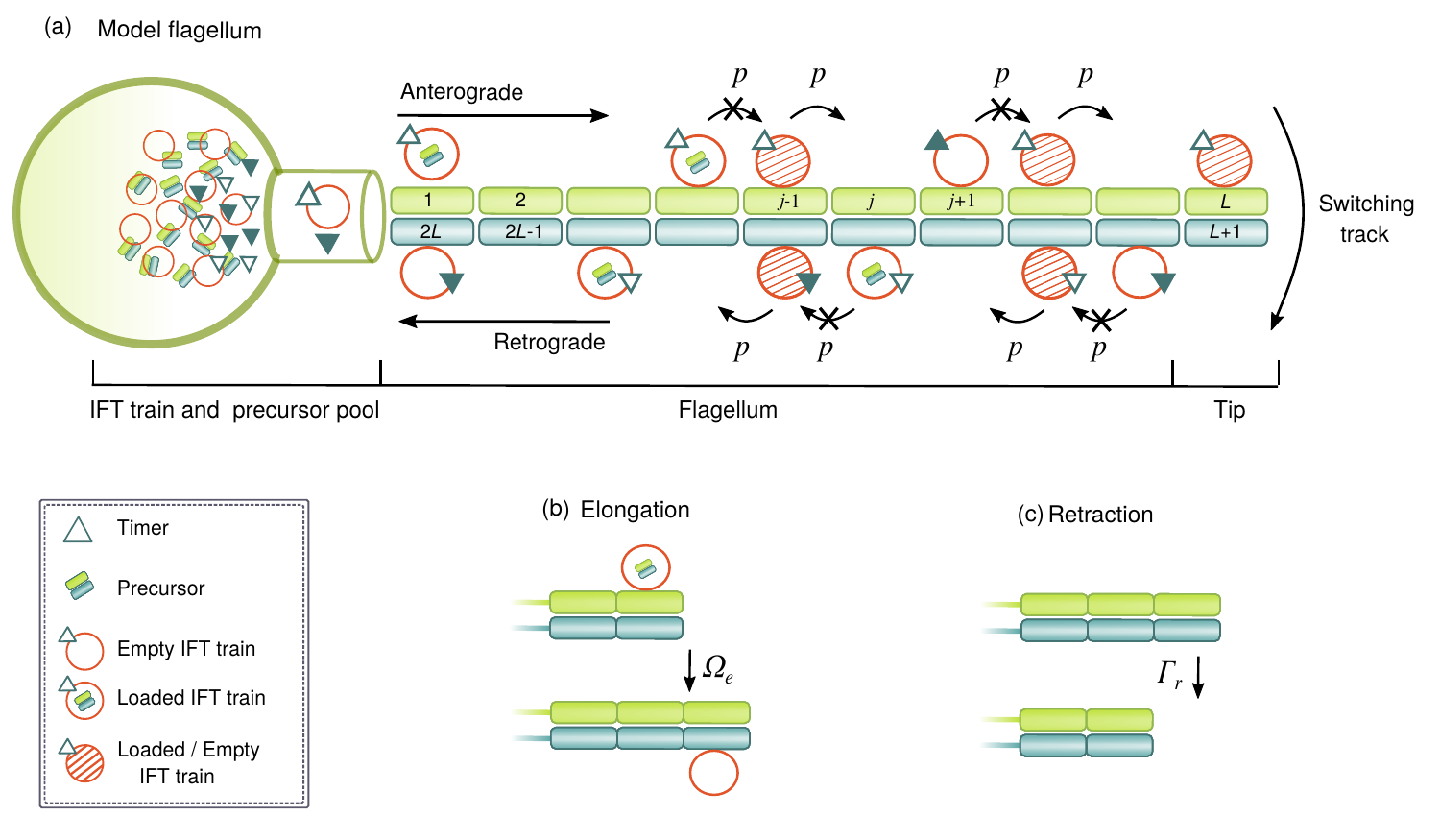}
\caption{ \textbf{Model :}   (a) The cell body is shown schematically as a sphere that accommodates the pool of the precursors (blue-green lattice units),  IFT trains (red balls)  and timer molecules (triangles). The flagellum is modelled by two antiparallel one-dimensional lattices of equal length that protrude outward from the cell body. While entering the flagellum, an  IFT train can be either empty (hollow balls) or loaded (balls with a lattice unit inside them) with a precursor. A red ball filled with red lines represents an IFT train which can be either empty or loaded. Whether or not to load a precursor on a train at the point of entry is decided by the state of the timer molecule that has just returned to the base after completing its round-trip journey inside the flagellum. On the green lattice, IFT trains move unidirectionally from the cell body towards the tip (called `anterograde'  movement) and on the blue lattice, IFT trains move unidirectionally from tip towards the cell body (called `retrograde' movement). In both the directions, the loaded as well as the empty IFT trains obey the exclusion principle, i.e., no site can be occupied by more than one train simultaneously.  The trains move with a steady average velocity $v$ that, because of the interactions arising from mutual exclusion, is density-dependent. The lattice sites are labelled by integer index $j$ ($j=1, 2, \dots, 2L$). An IFT train begins anterograde journey by hopping onto the site  $j=1$, then switches from anterograde to retrograde motion at the flagellar tip by hopping  from the site $j=L$ to the adjacent 
site $j=L+1$, if the target site is empty, and, finally, completes its round trip travel by exiting from $j=2L$. 
 (b) Before switching direction at the flagellar tip, a loaded IFT train can either (i) elongate the flagellum by adding a single lattice site to both the green and blue lattices, with probability $\Omega_e$, and return to the base empty, or (ii) return to the base carrying its undelivered cargo, without elongating the flagellum, with probability $1-\Omega_e$. (c) If there is no IFT train on the distal tips of both the green or blue lattices, the flagellum can shorten by the chipping of those two sites with the rate $\Gamma_r$. (Adapted from ref.\cite{patra20}.) }
\label{fig-model}
\end{figure}


As stated in the introduction, the ``precursors'' required for the growth of a flagellum at its tip and those degraded from the tip are transported in the opposite directions along two distinct tracks by molecular trains that are pulled by two different families of molecular motors called kinesin and dynein, respectively (see Fig.\ref{fig-model}(a)).  Kinesins are responsible for base to tip (anterograde) transport along a track while dyneins drive the transport in the opposite direction (retrograde transport) along a distinct parallel track. The IFT trains switch their direction of movement only at the tip of the flagellum. This bi-directional transport of precursors by the active motorized transport machinery in a eukaryotic flagellum is known as intraflagellar transport (IFT).  Interestingly, IFT is necessary even in a fully grown flagellum to maintain its length at a steady mean value by dynamic balancing of the rates of its ongoing growth and disassembly.

Our model of a flagellum consists of two antiparallel lattices, as shown schematically in  Fig.\ref{fig-model}(a). The precursors are cargoes that are loaded onto IFT trains which are  pulled by molecular motors along the tracks laid coaxially within the flagellum.  We do not describe the motion of the motors explicitly in the model. Each IFT train is represented by a self-driven particle. The collective traffic-like movement of the IFT trains along the respective tracks are represented as a TASEP. On the lattice, an IFT train moves with an average velocity $v$ that is determined by the dynamical phase of the TASEP which, in turn, depends on the rates of entry and exit of the IFT trains into and out of the flagellum at its base.  

An IFT particle entering the flagellum can either be empty or loaded with a single  precursor; the probability of loading a precursor onto the IFT train is dependent on the instantaneous population of the precursors in the pool as well as the current flagellar length. Information on the current flagellar length is conveyed to the flagellar base by the timer molecule just returning from a round-trip flight along the flagellum. This molecule has two chemical (or conformational) states $S_{\pm}$ and it makes a transition spontaneously from the state $S_{+}$ to $S_{-}$ with the rate $k$: $S_{+} \xrightarrow []{k} S_{-}$. A timer molecule that enters the flagellum while in the state $S_+$, after shuttling inside a flagellum of length $L_M(t)$ in the mother cell in time $t_{tof}$, can remain in state $S_+$ with probability $e^{-kt_{tof}}=e^{-2kL_M(t)/v}$ which is also the probability of loading a precursor onto the next IFT train.

The number density of the particles on the track and their flux are denoted by the symbols $\rho$ and $J$, respectively. So the flux of trains loaded with a precursor is given by $(N(t)/N_\text{max})e^{-2L(t)/v}J$. The precursors carried by these trains can either be utilised at the tip, with probability $\Omega_e$, for extending it by one unit. On the other hand, due to ongoing turnover of the tubulins at the tip, the flagellum can shrink by removing a subunit with rate $\Gamma_r$ provided both the sites at the tip are empty. So the overall disassembly rate is $\Gamma_r(1-\rho)^2$. Therefore, in the interphase the time-dependence of the length of the mother flagellum is given by the equation  
\begin{equation}
\frac{dL_\text{M}(t)}{dt}=\bigg{[}\frac{N_\text{M}(t)}{N_{max}}\bigg{]}  e^{ (-{2 k {L_\text{M}}(t)}/{v})} J \Omega_e -\Gamma_r (1-\rho )^2. 
\label{len-t}
\end{equation}
The primary quantities that characterize the IFT particle traffic in the steady state are (i) average particle density $\rho$, (ii) the average particle flux $J$ (and the corresponding average velocity $v$). These are also the the quantities  which enter directly in our model through the assembly and the disassembly rates (see equation (\ref{len-t})). An IFT particle needs tens of seconds to complete a round trip inside the flagellum whereas ciliogenesis requires a time of the order of tens of minutes. Because of these well separated time scales, we are justified in making the following assumtions: (a) that the flagellar length remains practically constant during the time of a single flight of the IFT particle, and (b) that, for each length of an flagellum,  the TASEP attains steady state so quickly that (time-indepedent) steady-state values can be used for the number density $\rho$, flux $J$ and velocity $v$. More specifically, for the traffic of IFT particles,  we choose values of steady-state values of $\rho$  that correspond to one of the three dynamical phases of TASEP and use the corresponding values of flux $J~[=q\rho(1-\rho)]$ and velocity $v~[=q(1-\rho)]$.

The pool plays a significant role in the life cycle of a flagellum as it supplies fresh precursors to a flagellum and receives precursors degraded from the tip of the flagellum. Suppose,  the precursors are synthesized and degraded with the rates $\omega^+$ and $\omega^-$, respectively. If $N_\text{max}$ is the maximum number of precursors which could be accommodated in the pool and currently has $N(t)$ number of precursors, the rate of fresh synthesis in our model is proportional to $N_M(t)/N_\text{max}$. Thus, in the interphase, the population dynamics of the precursors  in the pool is given by
\begin{eqnarray}
\frac{dN_\text{M}(t)}{dt}={\omega^+} \bigg{[}1-\frac{{N_\text{M}}(t)}{{N_\text{max}}}\bigg{]}-{\omega^-} {N_\text{M}}(t)-\frac{dL_\text{M}(t)}{dt}
\label{pool-t}
\end{eqnarray}
As shown explicitly in ref.\cite{patra20}, the rate equations (\ref{len-t}) and (\ref{pool-t}) can be derived by extracting the equations for the mean length and mean precursor population from the corresponding pair of master equations that capture the underlying  stochastic kinetic processes.

How should the Eq.(\ref{len-t}) and (\ref{pool-t}) be modified to account for the ciliary dynamics just before and after cell division? Instead of a single equation for $L_\text{M}(t)$ that describes the kinetics of the flagellum of the monoflagellate mother cell in its interphase, at least two appropriately modified equations for $L_\text{D1}(t)$ and $L_\text{D2}(t)$ are needed for describing the time-dependence of the lengths of the two flagella that the two daughter cells would inherit after cell division. As described above, in the interphase of the mother cell there exists a single pool that we refer to as the mother pool. But, just prior to cell division, this pool gets gradually divided into two separate pools that will be eventually inherited by the two daughter cells after cell division. So the population dynamics of the coupled pools of the mother and the two daughter cells require some appropriate extension of the equation (\ref{pool-t}). 

\begin{figure}
\includegraphics[scale=0.90]{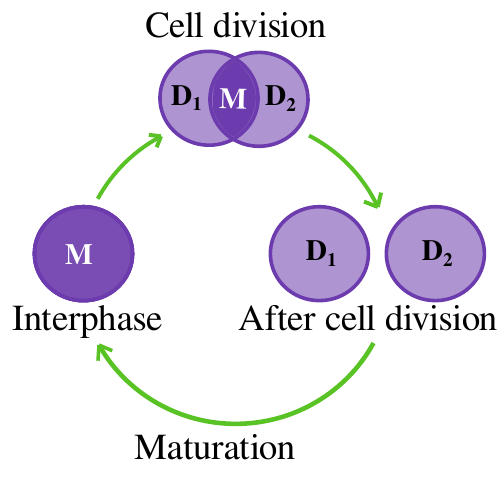}
\caption{ \textbf{Precursor pool and the cell cycle :} Mother pool, denoted by $M$, is the sole supplier of precursors during the interphase. Prior to the cell division two pools, denoted by $D_1$ and $D_2$, begin to form by the gradual transfer of precursors from the mother pool $M$. The pools $D_1$ and $D_2$ inherited by the two daughter cells mature to become the mother pool $M$ of the respective daughters in the interphase of their own lifetime. } 
\label{pool}
\end{figure}

The mother pool is schematically represented by $M$ in Fig.\ref{pool} and its population is denoted by $N_\text{M}$.  Just before cell division, two precursor pools, which will be inherited by the two daughter cells and are represented schematically by $D_1$ and $D_2$ in Fig.\ref{pool},  begin to form by gradual transfer of precursors from the mother pool with the rates $k_{cy1}$ and $k_{cy2}$, respectively. However, at any arbitrary instant during this phase, a fraction of the mother pool still remains in the overlap region (see Fig.\ref{pool}) although the mother cell would have stopped synthesis or degradation well before reaching this stage. 

Two assumptions are made at this stage: (i) Each daughter cell tunes the rate of her precursor synthesis on the premise that  the shrinking mother pool belongs to it, i.e.,  the total population of precursor of the $i$-th daughter is $N_\text{Di}+N_\text{M}$ ($i=1,2$). (ii) The freshly synthesized precursors get distributed among the daughter and the mother pools but each pool gets an amount proportional to its relative size. Similarly while degrading precursors, the amount degraded in each pool is assumed to be proportional to the precursors in each pool. Therefore, the equation governing the synthesis and degradation of precursors in the daughter-1 is obtained by quantifying
\begin{eqnarray}
\frac{d[N_\text{D1}(t)]}{dt}&=& \left( \substack{\text{Rate of synthesis} \\ \text{ in daughter 1} } \right)- \left( \substack{\text{Rate of degradation} \\ \text{ in daughter 1}} \right)
+ \left( \substack{ \text{Rate of transfer of precursors} \\ \text{ from  mother to daughter 1.}} \right)  
\label{eq-daughter1}
\end{eqnarray}
Similarly, the corresponding equation for the daughter-2 is obtained from (\ref{eq-daughter1}) by interchanging the symbols 1 and 2. The corresponding equation for the mother pool is obtained from 
\begin{eqnarray}
\frac{d[N_\text{M}(t)]}{dt}&=& \left( \substack{ \text{Rate of gain from synthesis} \\ \text{ in daughters 1 and 2} } \right) -\left( \substack{ \text{Rate of loss from degradation} \\ \text{ in daughters 1 and 2} } \right) 
- \left( \substack{ \text{Rate of transfer from  mother} \\ \text{ to  daughters 1 and 2.}} \right) 
\end{eqnarray}
The mathematical forms of these equations  are presented in the appendix (equation (\ref{mech-1-predivision}), (\ref{mech-1-division}) and (\ref{mech-2})). Solving these equations numerically, we get the temporal evolution of the mean flagellar lengths of the mother ($L_\text{M}(t)$) and the daughter flagella ($L_\text{D1}(t)$ and $L_\text{D2}(t)$) as well as the time-dependence of the mean precursor population in the respective pools ($N_\text{M}(t),~N_\text{D1}(t)$ and  $N_\text{D2}(t)$).

We also carry out the Monte Carlo (MC) simulations for monitoring the instantaneous flagellar length and precursor population for each realization of ciliogenesis which is essential for computing the correlations in the length fluctuations. The flagellar length  $L(t)$ takes discrete values $j=0,1,2,...$ and the pool population $N(t)$ can take values $n=0,1,2,..,N_\text{max}$. In a single MC time step, the instantaneous length $L(t)=j$ can either increase to $(j+1)$ or decrease to $(j-1)$. We calculate the probability of increasing or decreasing the length $L(t)=j$  from the effective rates of assembly $A_r=((N(t)/N_\text{max})e^{-2j/v}J \Omega_e$) and disassembly ($D_r=(1-\rho)^2 \Gamma_r$) respectively. The tip of the flagellum can be viewed as a non-homogenous random walker that moves towards its right $(j+1)$ with probability $A_r/(A_r+D_r)$ and towards its left $(j-1)$ with probability $D_r/(A_r+D_r)$. As the flagellum is coupled with precursor pools, the population of the pool is also updated when flagellar length changes. The length increases by removing a precursor from the pool and when the length decreases, one precursor is returned to the pool. During the pool separation, the probability of the flagellum exchanging precursors with a particular pool is given by the relative pool size (in terms of its population).  At every MC time step, the population is updated according to the rate of precursor synthesis ($Sy_r =\omega^+(1-j/M_\text{max})$)  and degradation ($De_r=\omega j$). The instantaneous precursor population in a given pool is increased by one with probability $Sy_r/(Sy_r+De_r)$ and decreased by one with probability $De_r/(Sy_r+De_r)$. 
The results obtained are presented and discussed in the next section of the main text. 

\section{Result}
We now present our results obtained by applying the formalisms developed above to the four species of monoflagellates \cite{he19,heimann89} that we described in section \ref{sec-organisms}.  In this section, we show how the mean flagellar length and the mean precursor population in the pool evolve with time as the cell enters different phases of the cell cycle. The results are obtained by numerically solving the equations  (\ref{mech-1-predivision}), (\ref{mech-1-division}) and (\ref{mech-2})) listed in the apppendix. 

\begin{figure}[h!]
\includegraphics[scale=1.1]{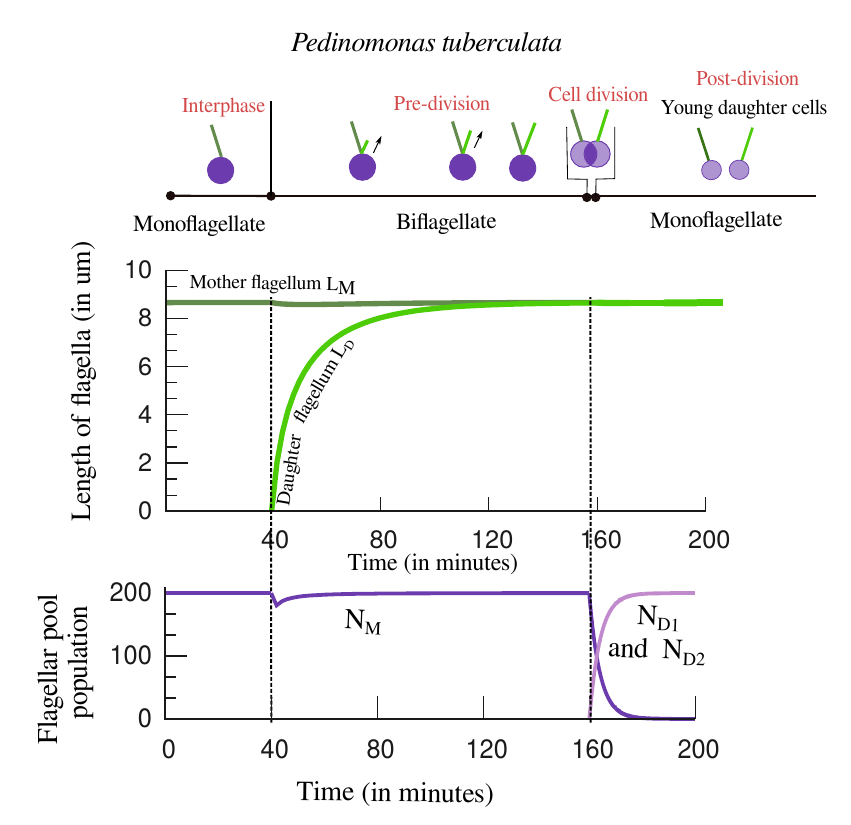}
\caption{\textbf{Mechanism I applied to ciliogenesis in \textit{Pedinomonas tuberculata }:} In the upper panel, the temporal evolution of the length of the daughter flagellum is seen while the length of the mother flagellum remains constant. The length of these two fully grown flagella remain unchanged while being distributed among the two daughter cells. In the lower panel, the evolution of the population of the mother pool and that of the daughter pools are shown during different phases of the cell cycle. ((Equation used : equation (\ref{mech-1-predivision}) and equation (\ref{mech-1-division}), Parameters :  $\omega^+=0.04,~\omega^-=1.0\times 10^{-4},~ N_\text{max}=400,~ k_\text{cy1}=k_\text{cy2}=5.0\times10^{-6},~ \Omega_e=0.2,~ \Gamma_r=1.0 \times 10^{-3},~ k=1.0\times 10^{-3},~J=0.09,~ \rho=0.1,~v=0.9, ~\delta t=4.0 \times 10^{-5} \text{minutes},~ \delta \ell =0.008 \mu m. $, Initial condition : ${N_M}(0)=200,{L_M}(0)=1100,{L_D}(0)=0$.}
\label{P-tuberculata}
\end{figure}

\begin{figure}[h!]
\includegraphics[scale=1.1]{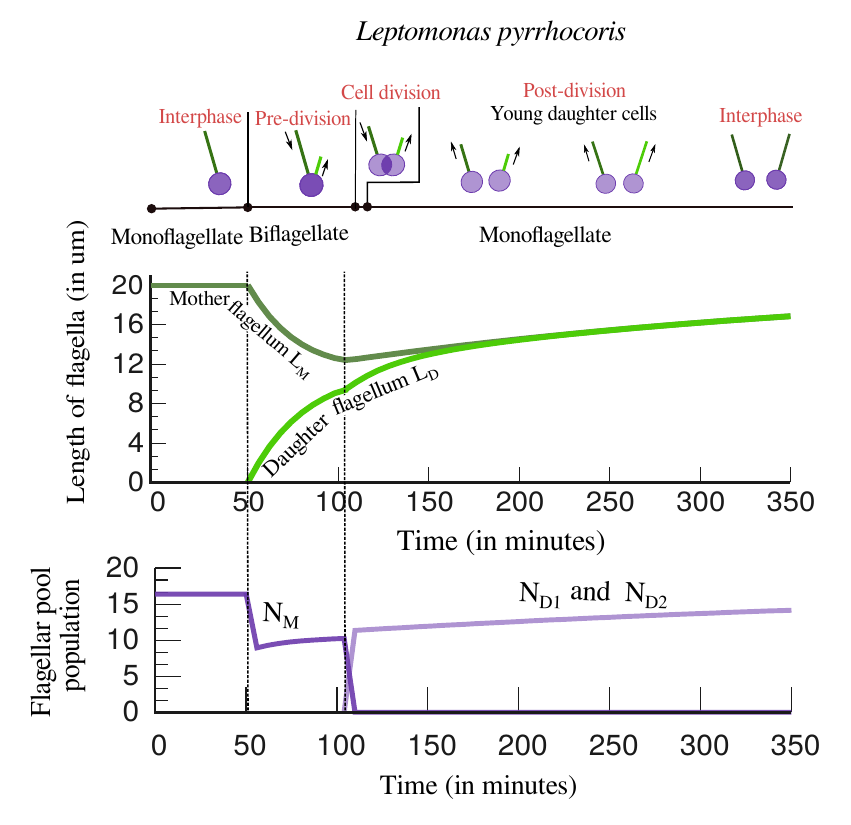}
\caption{\textbf{Mechanism I applied to ciliogenesis in \textit{Leptomonas pyrrhocoris}:} In the upper panel, the temporal evolution of the lengths of the mother and daughter flagella are shown. During the pre-division, the mother flagellum shortens and the daughter flagellum elongates simultaneously. After the cell division, both  flagella continue growing till they attain their interphase lengths. In the lower panel, the concomitant temporal variation of the population of the precursors in the mother pool and that in the daughter pools are shown during different phases of the cell cycle. (Equation used : equation (\ref{mech-1-predivision}) and equation (\ref{mech-1-division}), Parameters :  $\omega^+=2.0 \times 10^{-4},~\omega^-=1.0\times 10^{-5},~ N_\text{max}=100,~ k_\text{cy1}=k_\text{cy2}=5.0\times10^{-6},~ \Omega_e=0.33,~ \Gamma_r=2.0 \times 10^{-3},~ k=2.0\times 10^{-4},~J=0.09,~ \rho=0.1,~v=0.9, ~\delta t=2.0 \times 10^{-5} \text{minutes},~ \delta \ell =0.008 \mu m. $, Initial condition : ${N_M}(0)=16,{L_M}(0)=2500,{L_D}(0)=0$).}
\label{L-pyrrhocoris}
\end{figure}

\subsection{Mechanism I applied to {\it Pedinomonas tuberculata}}

By solving the set of coupled equations (\ref{len-t}) and (\ref{pool-t}), we first show that the mean length of the sole flagellum of {\it P. tuberculata} and the steady state population of the pool is maintained in the interphase (see Fig.\ref{P-tuberculata}). Then, by solving the set of coupled equations (\ref{mech-1-predivision}), we account for the flagellar replication in {\it P. tuberculata} in the pre-division phase as shown in the plot in Fig.\ref{P-tuberculata}. During this biflagellated phase, we managed to keep the mean length of the parental flagellum constant by choosing  sufficiently high rates of precursor synthesis ($\omega^+$). This high rate of precursor synthesis ensures that even during the ciliogenesis of the new flagellum, when precursors are delivered at a significant rate to the new flagellum , the pool still manages to have sufficient amount of precursors for the maintenance of the steady length of the parental flagellum. Once the daughter flagellum grows fully and becomes as long as the parental flagellum  $f_\text{CD} L^\text{(In)}_\text{M}$ $(f_\text{CD}=1)$, the bifurcation of the mother pool begins. We capture this phenomena using the set of the coupled equations mentioned in (\ref{mech-1-division}). Again due to the sufficiently high rate of synthesis of precursors in the pools to which the respective flagella are coupled, their lengths remain unchanged at this stage.

\subsection{Mechanism I applied to {\it Leptomonas pyrrhocoris}}
By solving the set of coupled equations (\ref{len-t}) and (\ref{pool-t}), we first show that the mean length of the sole flagellum of {\it L. pyrrhocoris} and the steady state population of the pool is maintained in the interphase (see Fig.\ref{L-pyrrhocoris}). Then, by solving the set of coupled equations (\ref{mech-1-predivision}), we account for the flagellar replication in {\it L. pyrrhocoris} in the pre-division phase as shown in the plot in Fig.\ref{L-pyrrhocoris}. During this biflagellated phase, the mother flagellum shortens. Due to the small population of the precursor in the pool and  slower rate of precursor synthesis ($\omega^+$), the cell could not meet the high demand of fresh precursors for the assembly of the new daughter flagellum. The longer mother flagellum continues releasing precursors due to ongoing turnover at the tip and send them back to the pool for recycling. However, due to greater demand, the recycled precursors are imported by the shorter daughter flagellum. As a result, the longer mother flagellum shortens because it is unable to import precursors to replace the loss caused by the turnover. The new daughter flagellum keeps growing at the expense of shortening of the mother flagellum. Cell division initiates when the daughter flagellum grows to $f_\text{CD}L^\text{(In)}_\text{M}=10 \mu m ~ (f_\text{CD}=0.5,~ L^\text{(In)}_\text{M}=20 \mu m)$ and the pool separation begins.  We capture these phenomena using the set of the coupled equations written in full detail in (\ref{mech-1-division}). At this point, due to length dependent incorporation of the precursors, the daughter flagellum incorporates precursors with a comparatively slower rate and  the precursor pool is also replenished. With the separation of pools, the mother flagellum finds a pool dedicated to it only and now could regrow back to its actual interphase length with the continuous supply of precursors from the pool which it no longer shares with the other growing daughter flagellum. Hence, the mean flagellar length of both the mother and the daughter flagella increase in a monotonous fashion in the post-division phase (see Fig.\ref{L-pyrrhocoris}). 

The phenomena of flagellar replication in monoflagellates, as described above, is very similar to the regeneration of a selectively amputated flagellum of a biflagellated {\it C. reinhardtii} whose other flagellum is left intact. The length of the intact flagellum remains practically unchanged while the amputated flagellum regrows to its full length provided the rate of precursor synthesis is sufficiently high \cite{patra20}. However, in the parameter regime that are physiologically realistic for {\it C. reinhardtii}, the rate of precursor synthesis is much lower. Consequently, during the regeneration of the amputated flagellum of {\it C. reinhardtii}, the longer flagellum shortens till the growing partner catches up and their lengths equalize; thereafter they grow together to achieve their steady state length. Very similar trend of variation, namely transient shortening, followed by elongation, of the parental flagellum can be observed also in the monoflagellate {\it P. tuberculata} if the rate of precursor synthesis is not sufficiently high. 

\subsection{Mechanism II applied to {\it Monomastix}}

\begin{figure}[h!]
\includegraphics[scale=0.8]{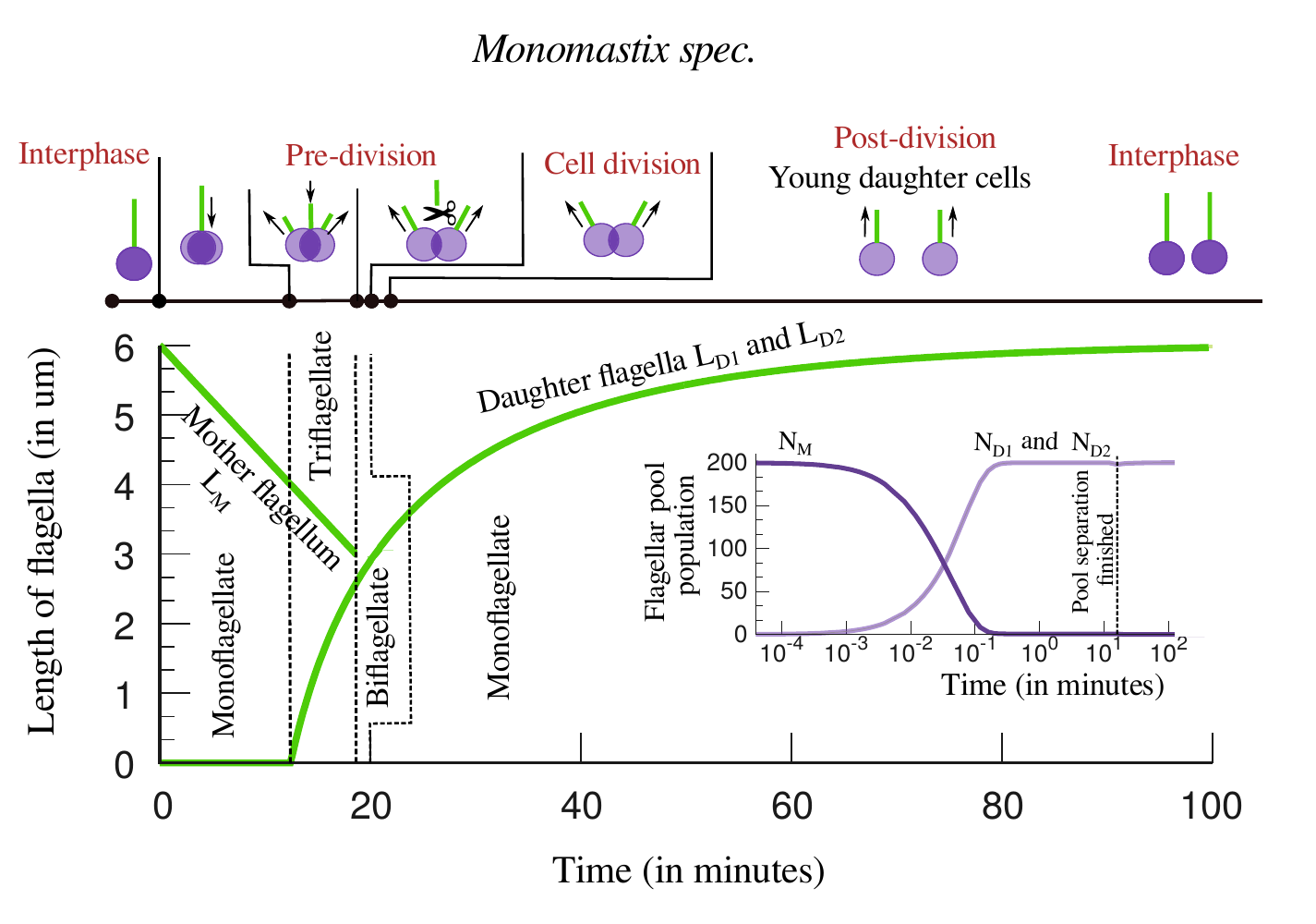}
\caption{ \textbf{Mechanism II applied to ciliogenesis in \textit{Monomastix spec} :} The time evolution of the length of the  shortening mother flagellum, which is later discarded by deflagellation, and the length of the elongating pair of  daughter flagella, which are inherited by the daughter cells, are shown. In the inset, the time evolution of the population of precursors in the mother pool and that in the two daughter pools are shown. (Equation used : equation (\ref{mech-2}), Parameters : 
$\omega^+=0.2,~\omega^-=8.0\times 10^{-4},~ N_\text{max}=1000,~ k_\text{cy1}=k_\text{cy2}=7.0\times10^{-4},~ \Omega_e=0.24,~ \Gamma_r=1.0 \times 10^{-3},~ k=1.0\times 10^{-3},~J=0.09 ~, \rho=0.1,~v=0.9, ~\delta t=4.0 \times 10^{-5},~\text {minutes},~ \delta \ell =0.008 ~ \mu m. $), Initial condition ${N_M}(0)=200,{N_{D1}}(0)=0,{N_{D2}}(0)=0,{L_M}(0)=750,{L_{D1}}(0)=0,{L_{D2}}(0)=0$.) } 
\label{Monomastix}
\end{figure}

We capture ciliogenesis in {\it Monomastix} using the set of coupled equations (\ref{mech-2}) formulated for the mechanism II and the solution is depicted in Fig.\ref{Monomastix}. In {\it Monomastix}, the pools begin to separate during the pre-division phase. During this phase, the parental flagellum starts to shorten (see Fig.\ref{Monomastix}). Pool separation results in monotonous decay in the population of precursors in the mother pool to which the parental flagellum is coupled  (see the inset in Fig.\ref{Monomastix}). As a result, the effective growth rate of the mother flagellum decreases monotonically to negligibly small value. The linear decay of mother flagellar length, observed in Fig.\ref{Monomastix} is a consequence of the overwhelming dominance of its disassembly rate $\Gamma_r (1-\rho)^2$. 

When the parental flagellum shrinks to length $f_\text{NF} L^\text{(In)}_\text{M}=4~\mu$m  ($f_\text{NF}=0.67, L^\text{(In)}_\text{M}=6 ~\mu$m), two new flagella emerge from the adjacent daughter pools. When the parental flagellum further shortens to length $f_\text{A} L^\text{(In)}_\text{M}=3~\mu$m ($f_\text{A}=0.5,~ L^\text{(In)}_\text{M}=6 ~\mu$m), it gets discarded by deflagellation. Before discarding the parental flagellum, the cell becomes tri-flagellated for a very brief period with two growing daughter flagella and one shortening parental flagellum. During this brief period, the daughter flagella grow quickly to  half of the normal interphase length. The amputation of the parental flagellum is followed by a stage in which the cell becomes effectively biflagellated. Thereafter, with the the complete separation of the pools, two new monoflagellated daughter cells emerge, each bearing a single  immature flagellum (in terms of length). The flagellum of each of these monoflagellate daughter cells keep growing till they attain their normal interphase length.  

\subsection{Mechanism II for {\it Pseudopedinella elastica}}
\begin{figure}[h!]
\includegraphics[scale=0.8]{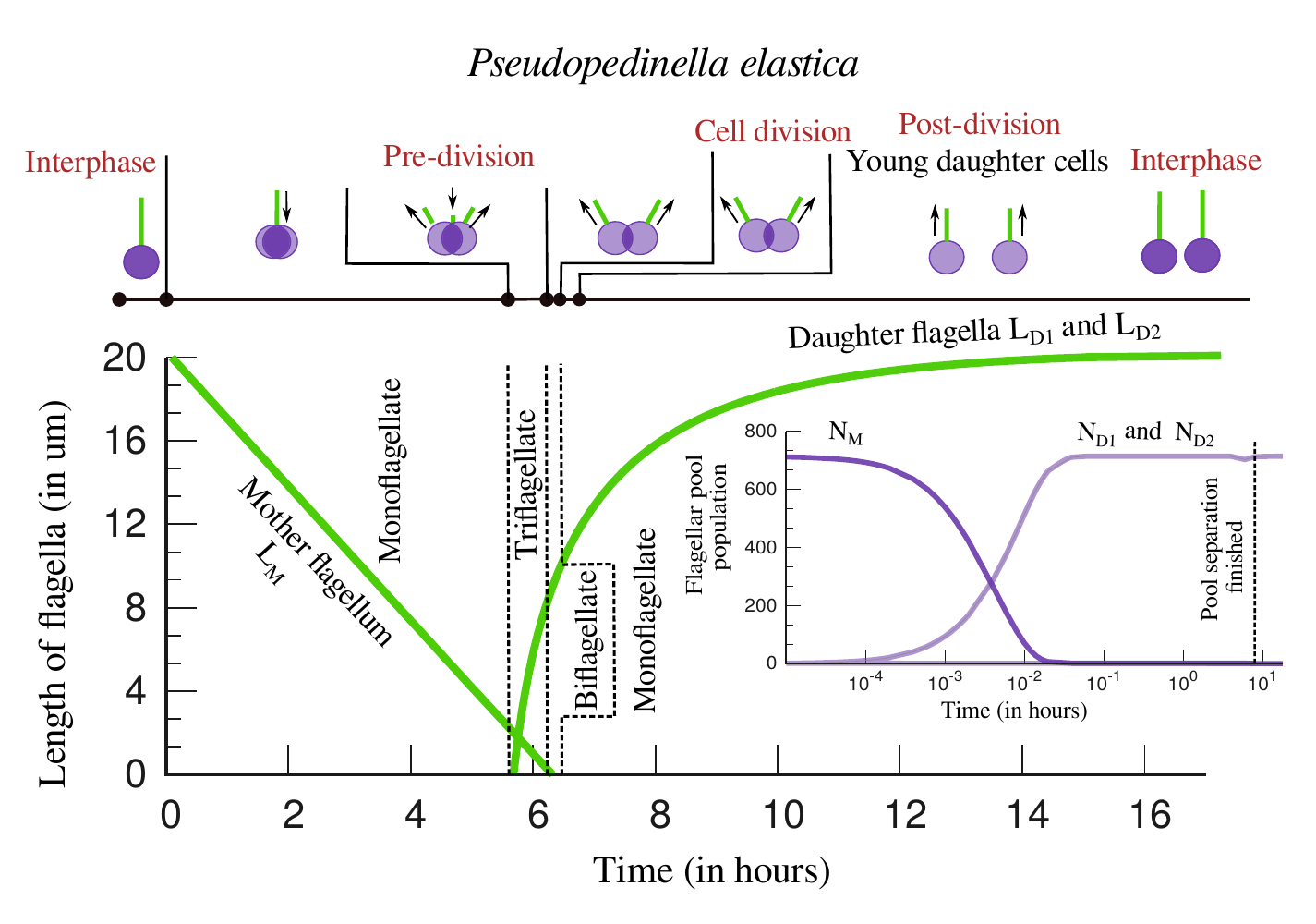}
\caption{ \textbf{Ciliogenesis in \textit{Pseudopedinella elastica} : }The time evolution of the length of the  shortening mother flagellum, which is fully resorbed, and the length of the elongating pair of  daughter flagella, which are inherited by the daughter cells, are shown. In the inset, the time evolution of population of precursors in the mother pool and that in the two daughter pools are shown. ((Equation used : equation (\ref{mech-2}) ,  Parameters: $\omega^+=0.2,~\omega^-=8.0\times 10^{-5},~ N_\text{max}=1000,~ k_\text{cy1}=k_\text{cy2}=3.0\times10^{-4},~ \Omega_e=0.12,~ \Gamma_r=1.0 \times 10^{-3},~ k=4.0\times 10^{-4},~J=0.09 ~, \rho=0.1, ~\delta t=2.0 \times 10^{-6}~ \text{hours} ~, \delta \ell =0.008~ \mu m. $ , Initial condition ${N_M}(0)=700,{N_{D1}}(0)=0,{N_{D2}}(0)=0,{L_M}(0)=2535,{L_{D1}}(0)=0,{L_{D2}}(0)=0$. )} 
\label{P-elastica}
\end{figure}

The quantitative description of cell division and ciliogenesis in {\it P. elastica} displayed in Fig.\ref{P-elastica} have been obtained by solving the set of coupled equations (\ref{mech-2}). 
Ciliogenesis in {\it P. elastica} is very similar to that in Monomastix except for three main features: (i) the length of the flagellum of {\it P. elastica} in the interphase is about 20 $\mu$m whereas that in {\it Monomastix} is about 6 $\mu$m, (ii) instead of the numerical value $f _\text{NF}$ = 0.67 in case of {\it Monomastix}, the corresponding value for {\it  P. elastica} is $f_\text{ NF}$ = 0.1 which implies that two new flagella emerge when the parental flagellum in {\it P. elastica} shortens to $f_\text{NF} L^\text{(In)}_\text{M}=2\mu$m ($f_\text{NF}=0.1, ~ L^\text{(In)}_\text{M}=20 \mu m)$ and (iii) in contrast to the numerical value $f_\text{A} = 0.5$ in case of {\it Monomastix}, the corresponding value for {\it P. elastica} is $f_\text{ A }$= 0 which implies that the parental flagellum completely resorbs. 
All the qualitative features of ciliogenesis in {\it P. elastica} are captured by solutions of our model equation as plotted in Fig.\ref{P-elastica}.

\section{Correlations in length fluctuations during multi-flagellated stages}

We need to monitor the instantaneous flagellar length in each realization of ciliogenesis for computing the correlation between the length fluctuations of distinct flagella of a single monoflagellate cell during its muti-flagellated transient existence. Hence, we carry out the Monte Carlo (MC) simulation of the model as described in section \ref{sec-model}. We generate 1000 sample trajectories for all the flagella which the cell bears in the multi-flagellated state and extract the instantaneous flagellar lengths for calculating the correlation function as described below .

We label the two distinct flagella of our interest by the indices $\mu$ and $\nu$. Before presenting the results, let us introduce the following definitions: suppose, the total number of realizations generated by the MC sampling is $n$. Let $L_{\mu}^i(t)$ and $L_{\nu}^i(t)$ denote the length of flagellum $\mu$ and $\nu$ at time $t$ in $i^{th}$ realization. The instantaneous {\it mean} length of the flagella are defined as 
\begin{equation}
\langle L_{\mu}(t) \rangle=\frac{\sum_{i=1}^n {L_{\mu}}^i(t)}{n}, ~{\rm and}~ \langle L_{\nu}(t) \rangle=\frac{\sum_{i=1}^n {L_{\nu}}^i(t)}{n},
\end{equation}
while the corresponding {\it variances} are given by
\begin{eqnarray}
Var(L_{\mu})= \bigg{[}\frac{1}{n-1}{\sum_{i=1}^n(\langle L_{\mu}(t) \rangle - {L_{\mu}}^i(t))^2}\bigg{]}^{1/2}, ~ {\rm and}~ Var(L_{\nu})= \bigg{[}\frac{1}{n-1}{\sum_{i=1}^n(\langle L_{\nu}(t) \rangle - {L_{\nu}}^i(t))^2}\bigg{]}^{1/2}.  \nonumber \\
\end{eqnarray}
and the \textit{covariance} $Cov(L_{\mu}L_{\nu})$ is given by
\begin{eqnarray}
\nonumber\\
\frac{1}{n-1}\bigg{[}{\sum_{i=1}^n(\langle L_{\mu}(t) \rangle - {L_{\mu}}^i(t))(\langle L_{\nu}(t) \rangle - {L_{\nu}}^i(t))}\bigg{]}^{1/2}
\end{eqnarray}
In terms of these {\it variances} and the {\it covariance}, 
the \textit{correlation} between the flagellar lengths is defined as 
\begin{eqnarray}
Corr(L_{\mu}L_{\nu})=\frac{Cov(L_{\mu}L_{\nu})}{Var(L_{\mu})Var(L_{\nu})};
\end{eqnarray}
and it gives a quantitative measure of the correlation of fluctuations in the lengths of the two flagella labelled by 
$\mu$ and $\nu$. 

In the above formulae, $t$ denotes the MC step. The averaging is carried out over large number of sample paths by repeating the process. The correlations provide insight into not only the nature of communication among the flagella, but also the extent of sharing of the the precursor pool among the flagella during the pre-division phase.  

\begin{figure*}
\includegraphics[scale=0.42]{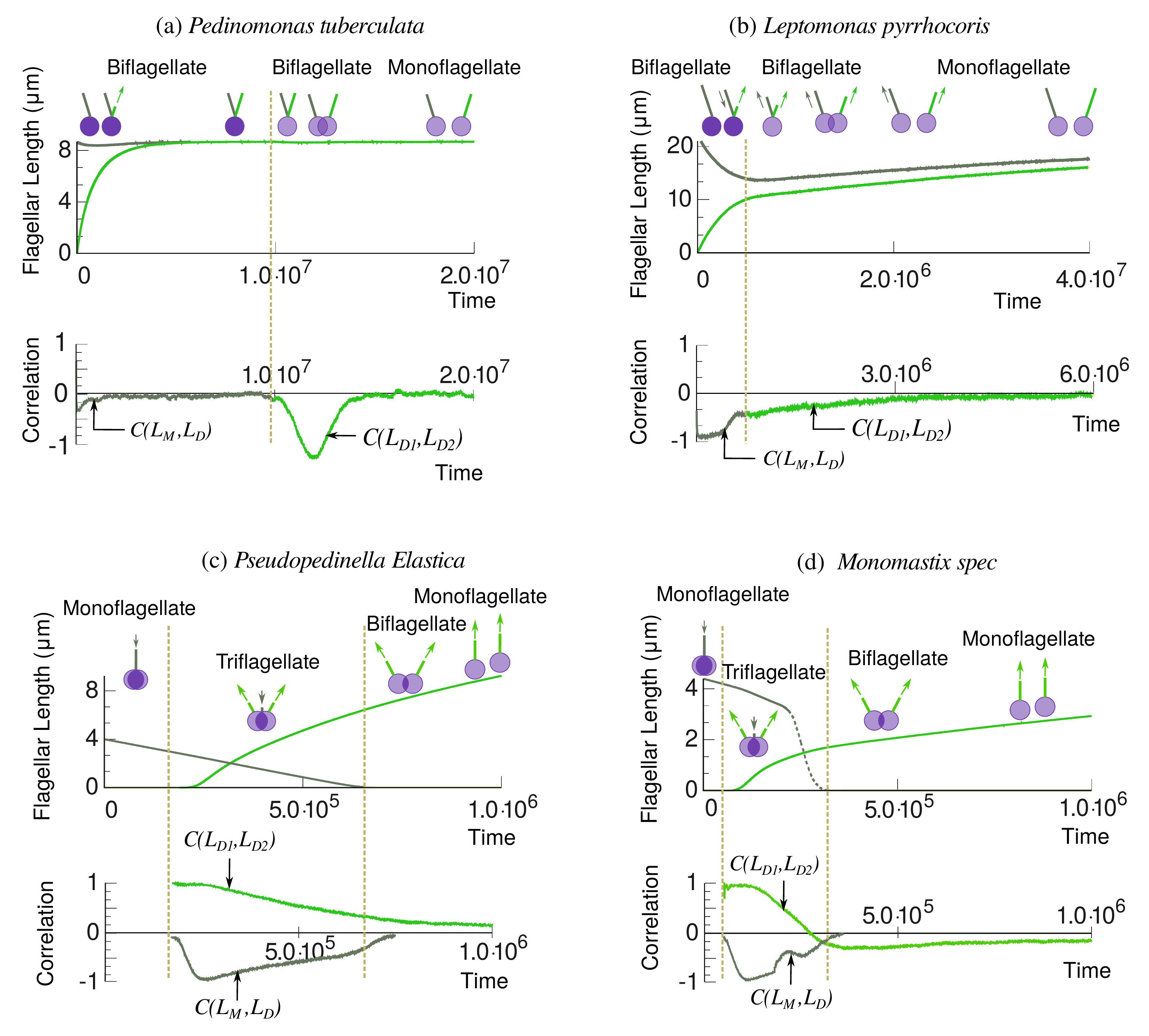}
\caption{ \textbf{Correlation : }  Correlation between the fluctuations of the lengths of two flagella of a single monoflagellates cell during its muti-flagellated transient existence. The parameters for the MC simulation of  (a) {\it Pedinomonas tuberculata}, (b) {\it Leptomonas pyrrhocoris} (c) {\it Monomastix spec} and (d){ \it Pseudopedinalla elastica} are listed in the caption of Fig.\ref{P-tuberculata}, Fig.\ref{L-pyrrhocoris} , Fig.\ref{P-elastica} respectively and Fig.\ref{Monomastix}.}
\label{corr}
\end{figure*}

\subsection{Correlations of flagellar length fluctuations in {\it Pedinomonas tuberculata}- Mechanism I}
\subsubsection{Correlation between length fluctuations of the mother and daughter flagella $C({L_M,L_D})$ in pre-division stage}
The correlation in fluctuation of length of the mother ($M$) and the new daughter flagellum ($D$) is denoted by $C(L_M,L_D)$. Initially, when the new daughter flagellum emerges and starts growing, $C(L_M,L_D)$ becomes slightly negative and thereafter becomes zero (see figure \ref{corr} (a)). Due to the high assembly rate of the daughter flagellum, initially the common mother pool starts to get depleted. So, the precursor released by the mother flagellum by ongoing turnover at its tip is utilised by the daughter flagellum for its own growth for a brief initial period making  $C(L_M,L_D)$ negative. In the mean time, the pool gets replenished by the synthesis of fresh  precursors, and therefore, thereafter the length fluctuations become uncorrelated. 

\subsubsection{Correlation between fluctuations of length of the  two daughter flagella $C({L_{D1},L_{D2}})$ in cell division stage}
During cell division, we refer the mother flagellum $M$ as $D_1$ and the new daughter flagellum $D$ by $D_2$. Now the daughter pools begin  to emerge from the mother pool and, consequently, the population of mother pool starts to fall due to transfer of precursors to the daughter pools. At this stage the daughter pools don't have sufficient precursors. So the correlation (now denoted by $C(L_{D1},L_{D2})$) becomes negative as growth of one flagellum takes place at the cost of shrinkage of the other. But ,once the daughter pools grow to sufficiently large size by synthesizing fresh precursors, the flagella start exchanging precursors with the respective daughter pools and the correlation among their length fluctuations start decreasing in magnitude and ultimately becomes zero indicating complete separation of the two daughter pools (see figure \ref{corr} (a)).

\subsection{Correlations of flagellar length fluctuations in {\it Leptomonas pyrrhocoris}- Mechanism I}
\subsubsection{Correlation between length fluctuations of the mother and daughter flagella $C({L_M,L_D})$ in pre-division stage}
As the new daughter flagellum emerges, the precursor pool immediately gets exhausted and the new flagellum grows by incorporating the precursors released by the mother flagellum, resulting in $C(L_M,L_D) \approx -1 $. This negative correlation indicates that the daughther flagellum grows at the cost of shortening the parental flagellum (see figure \ref{corr} (b)). 

\subsubsection{Correlation between fluctuations of length of the  two daughter flagella $C({L_{D1},L_{D2}})$ in cell division stage}
As the pool separation begins, two new pools emerge each dedicated exclusively to a single flagellum and the flagella start exchanging precursors with the pools they are attached with. Therefore, the magnitude of the correlation (now denoted by $C(L_{D1},L_{D2})$) decreases as the precursor pools segregate, and ultimately  becomes zero indicating complete separation of the precursor pools of the two daughter cells (see figure \ref{corr} (b)).

\subsection{Correlations of flagellar length fluctuations in {\it Pseudopedinella elastica} - Mechanism II}
\subsubsection{Correlation between length fluctuations of the mother and daughter flagellum $C({L_M,L_D})$}
In case of \textit{Pseudopedinella elastica}, interesting nontrivial behavior of the correlations functions are observed when the cell is effectively a transient triflagellate. The negative correlation function $C(L_M,L_D)$ (see figure \ref{corr} (b)) indicates shortening of the original flagellum of the mother and the concomitant elongation of the new flagella for the daughters.

\subsubsection{Correlation in fluctuations of the length of the daughter flagella $C({L_{D1},L_{D2}})$}

$C({L_{D1},L_{D2}})$ is initially positive and is of maximum magnitude because both the daughter flagella  start growing.  Then, as the flagella grow further, the correlation starts to decay with time and the fluctuations of the two become uncorrelated in {\it Pseudopedinella} as the separations of the two daughter cells near completion  (see figure \ref{corr} (b)).

\subsection{Correlations of flagellar length fluctuations in {\it Monomastix spec}- Mechanism II}
\subsubsection{Correlation between length fluctuations of the mother and daughter flagellum $C({L_M,L_D})$}

The correlation functions $C({L_M,L_D})$ and $C({L_{D1},L_{D2}})$ for {\it Monomastix spec} are qualitatively similar to those for {\it Pseudopedinella elastica} as both of these follow the mechanism -II. The minor quantitative difference between the correlation functions of the two monoflagellates arise from the fact that $f_\text{A}=0$ for one ({\it Pseudopedinella elastica}) while it is non-zero for the other ({\it Monomastix spec}).  

\section{Conclusion}

In this paper we have developed theoretical models for the length control of flagellum, which is a specific type of long cell appendage in eukaryotic cells. Specific systems that we have studied here are {\it monoflagellates}, which are uni-cellular organisms that possess a single flagellum for almost the entire period of their lifetime except a brief duration before their division into two daughter cells. A unique feature of these organisms is that for a short period just before their division they become multiflagellated.  

Flagellar length is regulated by bi-directional motorized transport of precursor proteins between the base and the distal tip of the flagellum. Our models combine the concepts of (i) {\it time-of-flight} mechanism for length sensing, (ii) {\it length-dependent loading} of precursor as cargo on intra-flagellar transport (IFT) trains pulled my molecular motors, and (iii) traffic of the IFT trains that are physical realizations of totally asymmetric simple exclusion process (TASEP).  Results of our model faithfully describe the time-dependence of the flagellar lengths. Note that we have not discussed here the implications which emerge due to the treatment of  IFT train traffic as TASEP. Other interesting features of the TASEP based flagellar  length control model are discussed in detail in our previous publication \cite{patra20}.

During the brief period when the monoflagellate cell leads the life of a multi-flagellate, the flagella communicate via the common pool of precursors that they share. In order to probe the nature of the cooperativity of the flagella that arise from such communication, we have computed correlations between the fluctuations of the lengths of different flagella by Monte Carlo (MC) simulations. A similar approach may be used in future for experimental study of the role of the common pool in terms of correlation functions.

\begin{appendix}
\section{Rate equations for ciliary length dynamics in monoflagellates}        
In this appendix we write down the rate equations that govern the ciliary dynamics in monoflagellates by mechanisms I and II described in the main text. 

\subsection{Mechanism I}
The coupled differential equations depicting their temporal evolution is:
\begin{subequations}
\begin{equation}
\frac{dL_\text{M}(t)}{dt}=\bigg{[}\frac{N_\text{M}(t)}{N_\text{max}}\bigg{]}  e^{ (-{2 k {L_\text{M}}(t)}/{v})} J \Omega_e -\Gamma_r (1-\rho )^2
\end{equation}
\begin{equation}
\frac{dL_\text{D}(t)}{dt}=\bigg{[}\frac{N_\text{M}(t)}{N_\text{max}}\bigg{]}  e^{ (-{2 k {L_\text{D}}(t)}/{v})} J \Omega_e -\Gamma_r (1-\rho )^2
\end{equation}
\begin{equation}
\frac{dN_\text{M}(t)}{dt}={\omega^+} \bigg{[}1-\frac{{N_\text{M}}(t)}{{N_\text{max}}}\bigg{]}-{\omega^-} {N_\text{M}}(t)-\frac{dL_\text{M}(t)}{dt}-\frac{dL_\text{D}(t)}{dt}
\end{equation}
\label{mech-1-predivision}
\end{subequations}

When $L_\text{D}(t)=f_\text{CD} L^\text{(In)}_\text{M}$, the cells begin to divide. The mother flagellum will be inherited by one progeny whereas the daughter flagellum will be inherited by another progeny. So, let us now change the notation $M \to D_1$ and $D \to D_2$. With this the coupled equations now become:
\begin{widetext}
\begin{subequations}
\begin{eqnarray}
\frac{d[L_\text{D1}(t)]}{dt} &=&\underbrace{\bigg{[} \frac{N_\text{D1}(t) + N_\text{M}(t)}{N_\text{max}} \bigg{]}e^{-2 k L_\text{D1}(t)/v}J\Omega_e - 
 \Gamma_r (1 - \rho)^2}_{\substack{\text{Daughter flagellum exchanges precursor with the mother and daughter pool}}}
\end{eqnarray}
\begin{eqnarray}
\frac{d[L_\text{D2}(t)]}{dt} &=&\underbrace{\bigg{[} \frac{N_\text{D2}(t) + N_\text{M}(t)}{N_\text{max}} \bigg{]}e^{-2 k L_\text{D2}(t)/v}J\Omega_e - 
 \Gamma_r (1 - \rho)^2}_{\substack{\text{Daughter flagellum exchanges precursor with the mother and daughter pool}}}
\end{eqnarray}
\begin{eqnarray}
\frac{d[N_\text{D1}(t)]}{dt}&=&\underbrace{\bigg{[}\frac{N_\text{D1}(t)}{N_\text{D1}(t)+N_\text{M}(t)}\bigg{]} \bigg{[} \omega^+ \bigg{[}1- \frac{N_\text{D1}(t)+N_\text{M}(t)}{N_\text{max}}\bigg{]} -\omega^-[N_\text{D1}(t)+N_\text{M}(t)]-\frac{d[L_\text{D1}(t)]}{dt}\bigg{]}}_{\substack{\text{Change in precursors population due to their synthesis,  degradation \& exchange with the flagellum. }}} \nonumber\\ &+&\underbrace{k_{cy1} N_\text{M}(t)\Theta(N_\text{M}(t))}_{\substack{\text{Precursor transferred from}\\ \text{mother pool till it exhausts.} }}
\end{eqnarray}
\begin{eqnarray}
\frac{d[N_\text{D2}(t)]}{dt}&=&\underbrace{\bigg{[}\frac{N_\text{D2}(t)}{N_\text{D2}(t)+N_\text{M}(t)}\bigg{]} \bigg{[} \omega^+ \bigg{[} 1-\frac{N_\text{D2}(t)+N_\text{M}(t)}{N_\text{max}}\bigg{]} -\omega^-[N_\text{D2}(t)+N_\text{M}(t)]-\frac{d[L_\text{D2}(t)]}{dt}\bigg{]}}_{\substack{\text{Change in precursors population due to their synthesis,  degradation \& exchange with the flagellum. }}}\nonumber\\ &+&\underbrace{k_\text{cy2} N_\text{M}(t)\Theta(N_\text{M}(t))}_{\substack{\text{Precursor transferred from}\\ \text{mother pool till it exhausts.} }}
\end{eqnarray}
\begin{eqnarray}
\frac{d[N_\text{M}(t)]}{dt}&=&\underbrace{\bigg{[}\frac{N_\text{M}(t)}{N_\text{D1}(t)+N_\text{M}(t)}\bigg{]} \bigg{[} \omega^+ \bigg{[}1- \frac{N_\text{D1}(t)+N_\text{M}(t)}{N_\text{max}}\bigg{]} -\omega^-[N_\text{D1}(t)+N_\text{M}(t)]-\frac{d[L_\text{D1}(t)]}{dt}\bigg{]}}_{\substack{\text{Change in precursors population due to their synthesis,  degradation \& exchange with the daughter flagellum $D1$. }}} \nonumber\\&+&\underbrace{\bigg{[}\frac{N_\text{M}(t)}{N_\text{D2}(t)+N_\text{M}(t)}\bigg{]} \bigg{[} \omega^+ \bigg{[} 1-\frac{N_\text{D2}(t)+N_\text{M}(t)}{N_\text{max}}\bigg{]} -\omega^-[N_\text{D2}(t)+N_\text{M}(t)]-\frac{d[L_\text{D2}(t)]}{dt}\bigg{]}}_{\substack{\text{Change in precursors population due to their synthesis,  degradation \& exchange with the daughter flagellum $D2$. }}} \nonumber\\ &-&\underbrace{\frac{d[L_\text{M}(t)]}{dt}}_{\substack{\text{Exchanging precursors with }\\ \text{the mother flagellum} }} } - \underbrace{ [k_\text{cy1}+k_\text{cy2} ]N_\text{M}(t) \Theta(N_\text{M}(t))}_{\substack{\text{Tranferring precursors to the} \\ \text{ daughter pools}}
\end{eqnarray}
\label{mech-1-division}
\end{subequations}
\end{widetext}

\subsection{Mechanism II}

 The set of coupled equations which describe the entire temporal evolution of the mother and the daughter flagella and the pools as well are given by:
\begin{widetext}
\begin{subequations}
\begin{eqnarray}
\frac{d[L_\text{D1}(t)]}{dt} =\underbrace{\bigg{[}\bigg{[} \frac{N_\text{D1}(t) + N_\text{M}(t)}{N_\text{max}} \bigg{]}e^{-2 k L_\text{D1}(t)/v}J\Omega_e - 
 \Gamma_r (1 - \rho)^2\bigg{]}}_{\substack{\text{Daughter flagellum exchanges precursor with the mother and daughter pool}}} \times \underbrace{\Theta(f_\text{NF}L^\text{(In)}_\text{M}-L_\text{M}(t))}_{\substack{\text{Ensures the elongation of daughter flagellum} \\ \text{ when mother shortens to $f_\text{NF}L^\text{(In)}_\text{M}$. }}}
\end{eqnarray}
\begin{eqnarray}
\frac{d[L_\text{D2}(t)]}{dt} =\underbrace{\bigg{[}\bigg{[} \frac{N_\text{D2}(t) + N_\text{M}(t)}{N_\text{max}} \bigg{]}e^{-2 k L_\text{D2}(t)/v}J\Omega_e - 
 \Gamma_r (1 - \rho)^2\bigg{]}} _{\substack{\text{Daughter flagellum exchanges precursor with the mother and daughter pool}}} \times \underbrace{\Theta(f_\text{NF}L^\text{(In)}_\text{M}-L_\text{M}(t))}_{\substack{\text{Ensures the elongation of daughter flagellum} \\ \text{ when mother shortens to $f_\text{NF}L^\text{(In)}_\text{M}$. }}}
\end{eqnarray}
\begin{equation}
\frac{d[L_\text{M}(t)]}{dt}= \underbrace{ \bigg{[} \frac{N_\text{M}(t)}{N_\text{max}} e^{-2 k L_\text{M}(t)/v}  J \Omega_e
  - \Gamma_r (1 - \rho)^2 \bigg{]}}_{\substack{\text{Mother flagellum exchanges precursor with the mother pool only}}} \times \underbrace{\Theta(L_\text{M}(t)-f_{A}L^\text{(In)}_\text{M} )}_{\substack{\text{Ensures that when the mother flagellum shortens to }\\ \text{ $f_{A}L^\text{(In)}_\text{M}$ and gets discarded, it stops changing its length} \\ \text{by exchanging precursors with mother pool}}}
\end{equation}
\begin{eqnarray}
\frac{d[N_\text{D1}(t)]}{dt}&=&\underbrace{\bigg{[}\frac{N_\text{D1}(t)}{N_\text{D1}(t)+N_\text{M}(t)}\bigg{]} \bigg{[} \omega^+ \bigg{[} 1-\frac{N_\text{D1}(t)+N_\text{M}(t)}{N_\text{max}}\bigg{]} -\omega^-[N_\text{D1}(t)+N_\text{M}(t)]-\frac{d[L_\text{D1}(t)]}{dt}\bigg{]}}_{\substack{\text{Change in population due to synthesis and degradation }\\ \text{of precursor by the daughter pool  is proportional to the relative size of the pool}  \\ \text{and by exchanging precursor with daughter flagellum}}} \nonumber\\ &+&\underbrace{k_\text{cy1} N_\text{M}(t)\Theta(N_\text{M}(t))}_{\substack{\text{Precursor transferred from}\\ \text{mother pool till it exhausts.} }}
\end{eqnarray}
\begin{eqnarray}
\frac{d[N_\text{D2}(t)]}{dt}&=&\underbrace{\bigg{[}\frac{N_\text{D2}(t)}{N_\text{D2}(t)+N_\text{M}(t)}\bigg{]} \bigg{[} \omega^+ \bigg{[}1- \frac{N_\text{D2}(t)+N_\text{M}(t)}{N_\text{max}}\bigg{]} -\omega^-[N_\text{D2}(t)+N_\text{M}(t)]-\frac{d[L_\text{D2}(t)]}{dt}\bigg{]}}_{\substack{\text{Change in population due to synthesis and degradation }\\ \text{of precursor by the daughter pool  is proportional to the relative size of the pool } \\ \text{and by exchanging precursor with daughter flagellum}}} \nonumber\\ &+&\underbrace{k_\text{cy2} N_\text{M}(t)\Theta(N_\text{M}(t))}_{\substack{\text{Precursor transferred from}\\ \text{mother pool till it exhausts.} }}
\end{eqnarray}
\begin{eqnarray}
\frac{d[N_\text{M}(t)]}{dt}&=&\underbrace{\bigg{[}\frac{N_\text{M}(t)}{N_\text{D1}(t)+N_\text{M}(t)}\bigg{]} \bigg{[} \omega^+ \bigg{[} 1-\frac{N_\text{D1}(t)+N_\text{M}(t)}{N_\text{max}}\bigg{]} -\omega^-[N_\text{D1}(t)+N_\text{M}(t)-\frac{d[L_\text{D1}(t)]}{dt}\bigg{]}}_{\substack{\text{Change in population due to synthesis and degradation }\\ \text{of precursor by the daughter pool  is proportional to the relative size of the pool.}}} \nonumber\\&+&\underbrace{\bigg{[}\frac{N_\text{M}(t)}{N_\text{D2}(t)+N_\text{M}(t)}\bigg{]} \bigg{[} \omega^+ \bigg{[} 1-\frac{N_\text{D2}(t)+N_\text{M}(t)}{N_\text{max}}\bigg{]} -\omega^-[N_\text{D2}(t)+N_\text{M}(t)]-\frac{d[L_\text{D2}(t)]}{dt}\bigg{]}}_{\substack{\text{Change in population due to synthesis and degradation }\\ \text{of precursor by the daughter pool  is proportional to the relative size of the pool.}}} \nonumber\\ &-&\underbrace{\frac{d[L_\text{M}(t)]}{dt}}_{\substack{\text{Exchange precursor with }\\ \text{the mother flagellum} }} } - \underbrace{ [k_\text{cy1}+k_\text{cy2} ]N_\text{M}(t) \Theta(N_\text{M}(t))}_{\substack{\text{Tranferring precursors to the} \\ \text{ daughter flagella during cell division}}
\end{eqnarray}
\label{mech-2}
\end{subequations}

\end{widetext}

\end{appendix}

{\bf Acknowledgement}: One of the authors (DC) acknowledges support from SERB (India) through a J.C. Bose National Fellowship. 

 
‌

\begin{thebibliography}{99} 

\bibitem{lee18} Robert Edward Lee (2018). {\it Phycology}. Cambridge Cambridge Univ Press.

\bibitem{marshall15b} Marshall W. F. (2015). Subcellular size. {\it Cold Spring Harbor perspectives in biology}, {\bf 7}(6), a019059. https://doi.org/10.1101/cshperspect.a019059

\bibitem{marshall15a} Marshall W. F. (2015). How Cells Measure Length on Subcellular Scales. {\it Trends in cell biology}, {\bf 25}(12), 760–768. https://doi.org/10.1016/j.tcb.2015.08.008

\bibitem{ludington15} Ludington, W. B. et al (2015). A systematic comparison of mathematical models for inherent measurement of ciliary length: how a cell can measure length and volume. {\it Biophysical journal}, {\bf 108}(6), 1361–1379. https://doi.org/10.1016/j.bpj.2014.12.051

\bibitem{patra20} Patra, S., J\"ulicher, F., Chowdhury, D. (2020). Flagellar length control in biflagellate eukaryotes: time-of-flight, shared pool, train traffic and cooperative phenomena.  {\it New Journal of Physics}, {\bf 22}(8), ab9ee4.  http://dx.doi.org/10.1088/1367-2630/ab9ee4 


\bibitem{ishikawa17} Ishikawa, H., and Marshall W., F., “Testing the time-of-flight model for flagellar length sensing.” {\it Molecular biology of the cell} {\bf 28},23 (2017): 3447-3456. doi:10.1091/mbc.E17-06-0384
‌
\bibitem{webb16} Webb, A. B., \& Oates, A. C. (2016). Timing by rhythms: Daily clocks and developmental rulers. {\it Development, growth \& differentiation}, {\bf 58}(1), 43–58. https://doi.org/10.1111/dgd.12242

\bibitem{lechtreck17} Lechtreck, Karl F et al. Protein transport in growing and steady-state cilia. {\it Traffic (Copenhagen, Denmark)} {\bf 18}, 5 (2017): 277-286. doi:10.1111/tra.12474

\bibitem{chowdhury13} D. Chowdhury, {\it Stochastic mechano-chemical kinetics of molecular motors: a multidisciplinary enterprise from a physicist's perspective}, Phys. Rep. {529}, 1 (2013).

\bibitem{kolomeisky15} A.B. Kolomeisky, {\it Motor proteins and molecular motors}, (CRC Press, 2015). 

\bibitem{chowdhury05} Chowdhury D., Schadschneider A., Nishinari K., (2005). Physics of Transport and Traffic Phenomena in Biology: from Molecular Motors and Cells to Organisms. {\it Physics of Life Reviews}, {\bf 2}(4), 318-352., doi:10.1016/j.plrev.2005.09.001. 

\bibitem{chowdhury00} Chowdhury D., Santen L., Schadschneider A., (2000). Statistical Physics of Vehicular Traffic and Some Related Systems. {\it Physics Reports}, {\bf 329},  199-329., doi:10.1016/s0370-1573(99)00117-9. 

\bibitem{schad10} A. Schadschneider, D. Chowdhury and K. Nishinari, {\it 
Stochastic transport in complex systems: from molecules to vehicles}, (Elsevier, 2010).

\bibitem{parmeggiani03}  A. Parmeggiani, T. Franosch, and E. Frey, Phys. Rev. Lett.
{\bf 90}, 086601 (2003); Phys. Rev. E {\bf 70}, 046101 (2004). 

\bibitem{nishinari05} K. Nishinari, Y. Okada, A. Schadschneider and
D. Chowdhury, {\it Intracellular transport of single-headed molecular
motors KIF1A}, Phys. Rev. Lett. {\bf 95}, 118101 (2005).

\bibitem{greulich07} P. Greulich, A. Garai, K. Nishinari, A. Schadschneider
and D. Chowdhury, {\it Intracellular transport by single-headed kinesin
KIF1A: effects of single-motor mechanochemistry and steric interactions},
Phys. Rev. E {\bf 75}, 041905 (2007).

\bibitem{chowdhury08b} D. Chowdhury, A. Garai and J.S. Wang, {\it Traffic
of single-headed motor proteins KIF1A: effects of lane changing},
Phys. Rev. E {\bf 77}, 050902 (R) (2008).


\bibitem{zia11}Zia, R. K. P., et al. (2011). Modeling Translation in Protein Synthesis with TASEP: A Tutorial and Recent Developments. {\it Journal of Statistical Physics}, {\bf 144}(2), 405-428. doi:10.1007/s10955-011-0183-1. 

\bibitem{chou11} T. Chou, K. Mallick and R.K.P. Zia, 
Rep. Prog. Phys. {\bf 74}, 116601 (2011).

\bibitem{johann12}Johann, D., Erlenk\"mper, C., \& Kruse, K. (2012). Length regulation of active biopolymers by molecular motors. {\it Physical review letters}, {\bf 108}(25), 258103. https://doi.org/10.1103/PhysRevLett.108.258103

\bibitem{melbinger12} Melbinger, A., Reese, L., \& Frey, E. (2012). Microtubule length regulation by molecular motors. {\t Physical review letters}, {\bf108}(25), 258104. https://doi.org/10.1103/PhysRevLett.108.258104

\bibitem{ghosh18} Ghosh S., et al. (2018) First-Passage Processes on a Filamentous Track in a Dense Traffic: Optimizing Diffusive Search for a Target in Crowding Conditions. {\it Journal of Statistical Mechanics: Theory and Experiment}, {\bf 12}, 123209. doi:10.1088/1742-5468/aaf31d. 

\bibitem{sharma12} Sharma A. K. \& Chowdhury D. (2012) Template-Directed Biopolymerization: Tape-Copying Turing Machines. {\it Biophysical Reviews and Letters}, {\bf 07}(03n04), 135-175., doi:10.1142/s1793048012300083. 

\bibitem{pinkoviezky13} Pinkoviezky I., \& Gov N. S., (2013). Modelling Interacting Molecular Motors with an Internal Degree of Freedom. {\it New Journal of Physics}, {\bf 15}(2) 025009., doi:10.1088/1367-2630/15/2/025009. 

\bibitem{sugden07} Sugden, K. E., Evans, M. R., Poon, W. C., \& Read, N. D. (2007). Model of hyphal tip growth involving microtubule-based transport. {\it Physical review. E}, {\bf 75} (3 Pt 1), 031909. https://doi.org/10.1103/PhysRevE.75.031909

\bibitem{schmitt11} Schmitt, M., \& Stark, H., (2011). Modelling bacterial flagellar growth. {\it Europhysics Letters}, {\bf 96}(2), 28001. doi:10.1209/0295-5075/96/28001

\bibitem{rolland15} Appert-Rolland, C., Ebbinghaus, M., \&  Santen, L. (2015). Intracellular transport driven by cytoskeletal motors: General mechanisms and defects. {\it Physics Reports}, {\bf 593}, 1-59. doi:10.1016/j.physrep.2015.07.001

\bibitem{sahoo16} Sahoo M., \& Klumpp S. (2016) Asymmetric Exclusion Process with a Dynamic Roadblock and Open Boundaries. {\it Journal of Physics A: Mathematical and Theoretical}, {\bf 49}(31) 315001., doi:10.1088/1751-8113/49/31/315001. 

\bibitem{john09} John A., et al. (2009). Trafficlike Collective Movement of Ants on Trails: Absence of a Jammed Phase. {\it Physical Review Letters}, {\bf 102}(10) doi:10.1103/physrevlett.102.108001. 

\bibitem{patra18} Patra S., \& Chowdhury D., (2018) Multispecies Exclusion Process with Fusion and Fission of Rods: A Model Inspired by Intraflagellar Transport. {\it Physical Review E}, {\bf 97}(1) doi:10.1103/physreve.97.012138. 

\bibitem{rosenbaum69} Rosenbaum, J.L., Moulder, J.E. and Ringo, D.L. (1969). Flagellar elongation and shortening in Chlamydomonas. {\it The Journal of Cell Biology}, {\bf 41}(2), pp.600–619.

\bibitem{marshall01} Marshall, W.F. and Rosenbaum, J.L. (2001). Intraflagellar transport balances continuous turnover of outer doublet microtubules. {\it Journal of Cell Biology}, {\bf 155}(3), pp.405-414.
‌\bibitem{ludington12} Ludington, W. B., Shi, L. Z., Zhu, Q., Berns, M. W., \& Marshall, W. F. (2012). Organelle size equalization by a constitutive process. {\it Current biology : CB,} {\bf 22}(22), 2173-2179. 

\bibitem{bauer20}Bauer, D., Ishikawa, H., Wemmer, K.A., Marshall, W.F. (2020). Analysis of Biological Noise in an Organelle Size Control System. bioRxiv 2020.08.31.276428; doi: https://doi.org/10.1101/2020.08.31.276428 


\bibitem{heimann89} Heimann, K., Benting, J., Timmermann, S. and Melkonian, M. (1989). The flagellar developmental cycle in uniflagellated algae. {\it Protoplasma}, {\bf 153}(1-2), pp.14-23.
. {\it Current Biology}, {\bf 22}(22), pp.2173–2179.

\bibitem{bertiaux18} Bertiaux, E. et al. “A Grow-and-Lock Model for the Control of Flagellum Length in Trypanosomes.” {\it Current biology : CB} {\bf 28},23 (2018): 3802-3814.e3. doi:10.1016/j.cub.2018.10.031


\bibitem{he19} He, C. Y., Singh, A., and Yurchenko, V., (2019). Cell Cycle-Dependent Flagellar Disassembly in a Firebug Trypanosomatid Leptomonas pyrrhocoris. {\it mBio} 10.6 (2019): e02424-19. 



\end{thebibliography}
\end{document}